\documentclass[tighten,preprint2,flushrt]{aastex}

\usepackage{epsfig,lscape} \usepackage[dvipdfm]{hyperref} \usepackage{natbib}
\newcommand{\myemail}{zouhu@nao.cas.cn}

\begin{document}

\title{Capability of Quasar Selection by Combining the SCUSS and SDSS Observations}
\author{Hu Zou \altaffilmark{1}, Xue-bing Wu \altaffilmark{2,3}, Xu Zhou \altaffilmark{1}, Shu Wang \altaffilmark{2}, Linhua Jiang\altaffilmark{3}, Xiaohui Fan \altaffilmark{4,3}, Zhou Fan \altaffilmark{1}, Zhaoji Jiang \altaffilmark{1}, Yipeng Jing \altaffilmark{5}, Michael Lesser \altaffilmark{4}, Cheng Li \altaffilmark{6}, Jun Ma \altaffilmark{1}, Jundan Nie \altaffilmark{1}, Shiyin Shen \altaffilmark{6}, Jiali Wang \altaffilmark{1}, Zhenyu Wu \altaffilmark{1}, Tianmeng Zhang \altaffilmark{1}, Zhimin Zhou \altaffilmark{1}}
\altaffiltext{1}{Key Laboratory of Optical Astronomy, National Astronomical Observatories, Chinese Academy of Sciences, Beijing, 100012, China; \myemail}
\altaffiltext{2}{Department of Astronomy, Physics School, Peking University, Beijing 100871, China}
\altaffiltext{3}{Kavli Institute for Astronomy and Astrophysics, Peking University, Beijing 100871, China}
\altaffiltext{4}{Steward Observatory, University of Arizona, Tucson, AZ 85721}
\altaffiltext{5}{Center for Astronomy and Astrophysics, Department of Physics and Astronomy, Shanghai Jiao Tong University, Shanghai 200240, China}
\altaffiltext{6}{Shanghai Astronomical Observatory, Chinese Academy of Sciences, Shanghai 200030, China}

\begin{abstract} 
The South Galactic Cap $u$-band Sky Survey (SCUSS) provides a 
deep $u$-band imaging of about 5000 deg$^2$ in south Galactic cap. It 
is about 1.5 mag deeper than the SDSS $u$-band. In this paper we 
evaluate the capability of quasar selection using both SCUSS and SDSS 
data, based on considerations of the deep SCUSS $u$-band imaging and 
two-epoch $u$-band variability. We find that the combination of 
the SCUSS $u$-band and the SDSS $griz$ band allows us to select more 
faint quasars and more quasars at redshift around 2.2 than the selection 
only with the SDSS $ugriz$ data. Quasars have significant $u$-band 
variabilities. The fraction of quasars with large two-epoch variability 
is much higher than that of stars. The selection by variability can 
select both low-redshift quasars with ultraviolet excess and mid-redshift 
($2 < z <3.5$) quasars where quasar selection by optical colors is 
inefficient. The above two selections are complementary and make full 
use of the SCUSS u-band advantages.
\end{abstract}



\section{Introduction}
Quasars, identified as high-redshift objects, are very distant according 
to the Hubble's law. The absorption spectra of quasars close to the 
reionization epoch are the best probes to study the intergalactic medium 
(IGM) and reionization \citep{fan02}. In recent years, large-scale optical 
surveys such as the Sloan Digital Sky Survey \citep[SDSS;][]{yor00} have 
provided numerous and homogeneous data for studying the properties of 
quasars and other subjects including connections to their host 
galaxies, IGM, Ly$\alpha$ forests, quasar clustering, central supermassive 
black holes, and baryon acoustic oscillations \citep{dun03,fan06,ros09,she09,daw13}, etc.

Quasar targets for spectroscopic observations in the SDSS are mainly 
selected by optical colors \citep{ric02}. Outliers away from 
the star sequence in color space are considered as spectroscopic 
follow-up candidates. Due to strong ultraviolet excess and emission lines, 
quasars with redshift lower than 2.2 or higher than 3.0 are obviously 
far from the star locus in color-color diagrams. A lack of or not so deep
$u$ band would significantly impact the ability to select the low-redshift 
quasars. Optical colors of quasars in the redshift range of 
$2.2 < z < 3.0$ are similar to those of A-colored stars (blue horizontal 
branch stars and blue stragglers) so that it is very difficult 
and inefficient to select quasars only by simple color cuts 
\citep{fan99,ric02,wux11}. There are two alternatives that can help to 
pick out those quasar targets more efficiently. One is combining optical 
and near-infrared (NIR) photometry, e.g. the Two Micron All Sky Survey 
\citep[2MASS;][]{skr06}, UKIRT Infrared Deep Sky Survey 
\citep[UKIDSS;][]{law07}, and Wide-field Infrared Survey Explorer 
\citep[WISE;][]{wri10}. The decreasing rate of a stellar continuum from 
optical to NIR wavelength is larger than that of a quasar. Optical 
colors together with infrared colors are utilized to distinguish 
quasars at $z > 2.2$ \citep{mad08,wux11,wux12}. 

The optical variability is the other way to select quasar candidates. 
Its role will become more and more prominent with the ongoing or 
up-coming time-domain surveys such as the Palomar Transient Factory 
\citep[PTF;][]{law09}, Panoramic Survey Telescope \& Rapid Response System 
\citep[PAN-STARRS;][]{kai04}, and Large Synoptic Survey Telescope 
\citep[LSST;][]{lss09}. Light curves have been parameterized by a structure 
function or a damped random walk model \citep{sch10,mac11,but11,pal11}. 
Quasars, variables and non-variable stars can be successfully 
separated with a great completeness and purity in the parameter space.
However, quasar targeting by light curves needs tens of time sampling 
points, and only limited sky areas have deep time-series photometric 
observations. The $u$-band variability is the largest among the five 
SDSS bands and it becomes larger as the time lag is longer. The 
magnitude change in $u$ band might be helpful for quasar targeting 
selection. 

The SCUSS is an imaging survey in the south Galactic cap (SGP) 
with an SDSS-like $u$ filter\citep{zho14}. The survey is about 1.5 mag 
deeper than the SDSS $u$ band. It is expected that, comparing with target 
selections only by SDSS photometry, one can select more quasars with less 
star contamination by using both deep SCUSS $u$ band and SDSS other bands 
if objects present no variability. The SCUSS observations started in 
2010 and ended in 2013. The average time lag between SCUSS and SDSS 
observations is about 2--3 years. We can obtain the two-epoch 
$u$-band variability between these two surveys. The variability difference 
between quasars and stars is useful for quasar selection.
In this paper, we compare the quasar selection by combining the SCUSS 
and SDSS data with that only by the SDSS data and try to understand the 
quasar targeting potential and capability when the SCUSS $u$ band is 
involved. In Section \ref{sec2}, we give brief descriptions of the SCUSS 
and our quasar and star samples. Section \ref{sec3} presents some advantages 
of the SCUSS $u$-band photometric data, which are favourable for the 
quasar selection. Quasar selections by combining SCUSS and SDSS data are 
analyzed and discussed in Section \ref{sec4}. Section \ref{sec5} is the 
conclusion. 

\section{Data} \label{sec2}
\subsection{SCUSS data}
The SCUSS is undertaken by the National Astronomical Observatories of China. 
The adopted telescope is the 2.3 m Bok telescope located on Kitt Peak, Arizona. 
The camera deployed at the prime focus provides a field of view of about 1 
square degree. The photometric system is the SDSS-like $u$ filter, whose 
effective wavelength is about 3538 {\AA} and FWHM is about 520 {\AA} 
\citep{zou14}. The filter is a little bluer than the SDSS $u$ band. The 
SCUSS covers an area of about 5000 square degrees in the SGP. The exposure 
time is 5 minutes and the 5-$\sigma$ magnitude limit is deeper than 23.0 mag. 

The detailed image processing and photometry for the SCUSS can be referred 
to the paper of \citet{zou14}. In general, the global astrometric accuracy 
is about 0.13{\arcsec}. The catalogs include both photometry for stacked 
images and co-added photometry for single-epoch images. The co-added PSF 
magnitudes and co-added aperture magnitudes, if a proper aperture radius 
is chosen, are the best brightness measurements of point sources. In this 
paper, we use the co-added PSF magnitude. The SCUSS magnitude is converted 
to the SDSS photometric system using the transformation equation of 
\begin{equation} 
u_\mathrm{SDSS}=u_\mathrm{SCUSS}+0.0586(u_\mathrm{SCUSS}-g)-0.0207(u_\mathrm{SCUSS}-g)^2-0.0377, \label{equ1}
\end{equation}
where $0.8< u_\mathrm{SCUSS}-g <2.8$. The corresponding error of the 
transformed SCUSS $u$-band magnitude is estimated by error transfer. This 
transformation is derived by point sources with SCUSS and SDSS photometric 
errors less than 0.05 mag. It is applied to point sources here and the 
maximum systematic difference between SCUSS and SDSS is about 0.036 mag. 
The SDSS magnitudes are PSF magnitudes. In the rest of this paper, 
all magnitudes are corrected by the Galactic extinction map of \citet{sch98}. 

\subsection{Quasar and star samples}
\begin{figure} \epsscale{1.0} \plotone{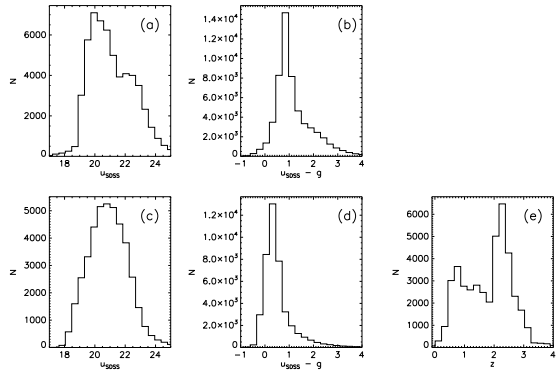} 
\caption{(a) and (b) are distributions of the $u_\mathrm{SDSS}$ magnitude and 
$u_\mathrm{SDSS} - g$ color of the star sample. (c), (d), and (e) are distributions of 
the $u_\mathrm{SDSS}$ magnitude, $u_\mathrm{SDSS} - g$ color, and 
redshift of the quasar sample, respectively.
\label{fig1}} 
\end{figure}

The spectroscopically confirmed quasars and stars are obtained by matching 
objects from the SDSS DR10 with SCUSS catalogs. We require that SCUSS 
objects are not saturated and not polluted by other sources and the SDSS 
classifications are point-like. The $i$-band magnitude is limited to the 
range between 18 and 22.5 mag. There are 42418 quasars and 
56518 stars in total. Figure \ref{fig1} shows some properties of these two 
samples including the distributions of the $u_\mathrm{SDSS}$ magnitude, 
$u_\mathrm{SDSS} - g$ color, and redshift (only for quasars). 
There are two peaks in the redshift distribution of quasars, 
locating in the ultraviolet-excess region and at intermediate redshift, 
respectively. Most mid-redshift quasars are obtained by the Baryon 
Oscillation Spectroscopic Survey \citep[BOSS;][]{daw13}, which plans to 
map the large-scale structure traced by the Ly-$\alpha$ forest. 

\section{SCUSS Advantages for Quasar Selection}\label{sec3}
\subsection{Deeper SCUSS photometry}
\begin{figure} \epsscale{1.0} \plotone{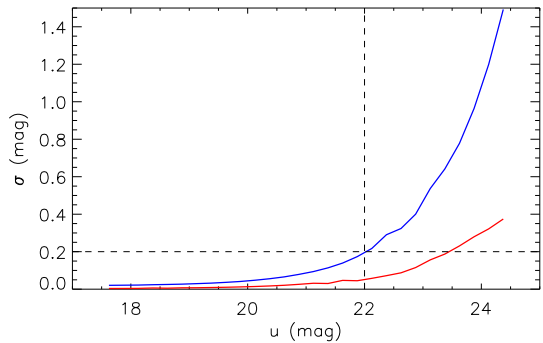}
\caption{Median magnitude error as a function of magnitude. 
The data come from a random SCUSS region with typical imaging depth. 
Here magnitudes are the ones without Galactic extinction corrections.
The blue curve is for the SDSS $u$ band, while the red one is for the 
SCUSS $u$-band. The dashed horizontal and vertical lines are corresponding 
to the photometric error of 0.2 mag and the magnitude of 22.0 mag.
\label{fig2}}
\end{figure}

The SCUSS $u$ band is reported to be 1.5 mag deeper than the SDSS 
$u$ band. We plot the magnitude error as a function of magnitude in Figure 
\ref{fig2}. The 5-$\sigma$ magnitude limits at $\sigma = 0.2$ for the SCUSS 
and SDSS $u$ bands are 23.45 and 22.03 mag, respectively. The magnitude 
error of the SDSS $u$-band at 22.0 mag, officially defined as the limiting 
magnitude of 95\% detection repeatability for point sources, is about 0.2 
mag , while that of the SCUSS $u$-band is about 0.05 mag.

\begin{figure} \epsscale{1.0} \plotone{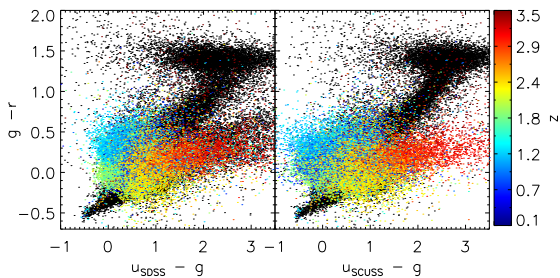}
\caption{Quasar and star samples in the $u_\mathrm{SDSS} - g$ and 
$u_\mathrm{SCUSS} - g$ vs. $g - r$ diagrams. Black dots are stars and points with colors 
are quasars colored by redshift. 
\label{fig3}}
\end{figure}

Another intuitive comparison of the data quality between these two surveys 
is to plot color-color diagrams and see the color dispersion of faint 
main-sequence stars. Figure \ref{fig3} shows the distributions of both quasar 
and star samples in the color-color plane of $u_\mathrm{SDSS} - g$ or 
$u_\mathrm{SCUSS} - g$ vs. $g - r$. The color dispersion of stars 
with SCUSS $u$ is obviously smaller than that with SDSS $u$. The 
$u_\mathrm{SCUSS} - g$ dispersion at $g - r = 1.4$ (mostly faint M stars) 
is about 0.55, while the $u_\mathrm{SDSS} - g$ dispersion is about 0.76. 
In addition, in the region occupied by quasars with $z > 2.5$, there are 
many stars initially selected as quasar candidates by the SDSS as shown 
in the left of Figure \ref{fig3}. They are mostly faint stars with large 
SDSS $u$-band photometric errors. However, most of these stars are still 
in the main sequence due to deeper SCUSS $u$-band as seen in the right 
panel of this figure. The deeper SCUSS $u$ band with SDSS other bands will 
evidently improve the quasar selection if quasars present no variability.  

\subsection{Two-epoch variability}
As seen in Figure \ref{fig3}, many quasars with $2 < z < 3$ are mixed 
with A-colored stars. The target selections only based on optical colors are 
inefficient for these quasars.  However, quasars usually present light 
variabilities. In principle, the variability would be larger 
as the wavelength is shorter and the observation lag is longer. The 
$u$-band variability is largest among all SDSS bands. There is a typical 
observation time lag of 2--3 years between the SCUSS and SDSS. Thus, 
the two-epoch $u$-band variability between these two surveys should be 
large enough as a useful tool to select quasar targets. The quasar selection 
by variability should also be complementary to other selection methods 
based on colors, especially for quasars with $2 < z < 3$.  

\begin{figure} \epsscale{1.0} \plotone{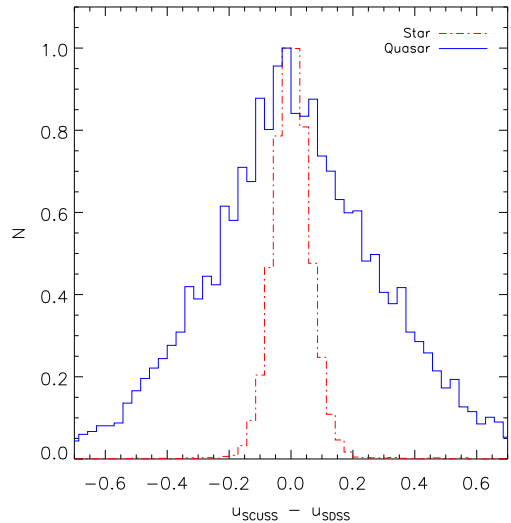}
\caption{Distributions of magnitude differences between the SCUSS and SDSS 
for quasars (solid line) and stars (dot-dashed line), which are normalized 
to their maximums. The $u_\mathrm{SCUSS}$ and $u_\mathrm{SDSS}$ magnitude 
errors are limited to be less than 0.05 mag.  
\label{fig4}}
\end{figure}

We compare the two-epoch magnitude differences of both stars and 
quasars in Figure \ref{fig4}. Only objects with photometric errors less than 
0.05 mag are considered. The standard deviations of the distributions for 
stars and quasars are about 0.06 mag and 0.3 mag, respectively. The scatter 
of stars mainly comes from the photometric error, while the scatter for 
quasars might come from the photometric error, intrinsic variability, and 
photometric system difference between the SCUSS and SDSS. Anyhow, 
the difference of the distributions between quasars and stars are distinct. 
It can be used to separate these two kinds of objects. The large two-epoch 
magnitude differences of quasars are also implied in the broader 
$u_\mathrm{SCUSS} - g$ color distributions in the right of Figure \ref{fig3}.

\subsection{Larger photometric system difference for quasars}
The SCUSS $u$ filter is very similar to the SDSS $u$ filter, but it is a little 
bluer. The central wavelength of the SCUSS $u$ band is about 3538 \AA, while 
that of the SDSS $u$ is about 3562 {\AA} \citep{zou14}. Stellar spectra 
are dominated by the continuum, so their photometric differences 
between the SCUSS and SDSS as shown in Equation (\ref{equ1}) is small. The 
photometric effect due to different photometric systems is less than 0.036 mag 
for main-sequence stars. However, quasar spectra present strong ultraviolet 
excess and emission lines. Their photometric differences should be much bigger. 

\begin{figure} \epsscale{1.0} 
\plotone{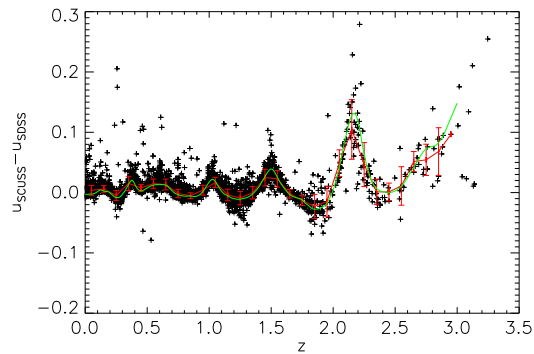}
\caption{Synthetic magnitude difference between the SCUSS and SDSS as 
a function of redshift. The magnitudes are calculated by convolving the 
quasar spectra blue-shifted by $z = 0.3$ in the SDSS DR10 with corresponding 
filter responses. The red solid line with error bars present the means and 
standard deviations in different redshift bins. The green dashed curve 
shows the magnitude difference derived by the composite quasar spectra. \label{fig5}}
\end{figure}

We get 4000 quasar spectra with different redshifts from the SDSS DR10. 
The wavelength of SDSS spectra ranges from 3800 to 9200 {\AA}, which is 
out of the $u$-band wavelength coverage. These quasar spectra are blue-shifted 
by $z = 0.3$ so that they can be convolved with two $u$ filter 
responses to generate synthetic magnitudes. Here, both SCUSS and SDSS 
filter responses are the ones with atmospheric extinction at the typical 
airmass of 1.3 \citep{zou14}. The resulting magnitude difference between 
these two filters is plotted as a function of redshift in Figure \ref{fig5}. 
In another way, we redshift the quasar composite spectrum by different 
redshift values and check the systematic variation with redshift as also 
shown in Figure \ref{fig5}. These two data sets present a coincident variation. 
Most quasars lie in a narrow band along the redshift. The wave-shape variation 
along the redshift is because of different emission lines entering and departing 
from the $u$ band. The magnitude difference is less than 0.03 when $z < 2.0$. 
It goes up to about 0.1 mag when $z > 2.0$, which is caused by the strongest 
Ly$\alpha$ line in quasar spectra. In contrast with stars, it is evident that 
quasars present much more difference induced by different photometric filters. 
This kind of difference is helpful for discriminating quasars from stars. Thus, 
we do not need to correct it and assume the magnitude difference mainly come 
from intrinsic variability if the photometric error is ignorable. 

\subsection{Quasar variability independent on redshift}
\begin{figure}
\epsscale{1.0}
\plotone{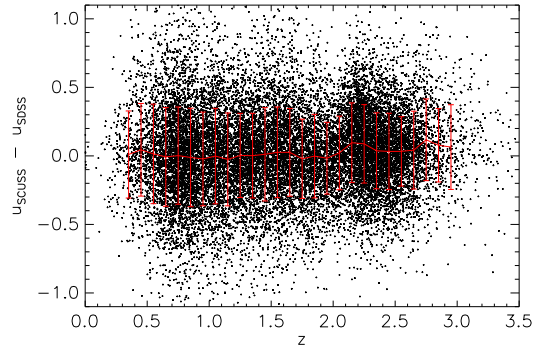}
\caption{Photometric magnitude differences of quasars between the SCUSS 
and SDSS as a function of the spectroscopic redshift. The overlapped red 
line with error bars show the averages with $1\sigma$ scatters in different 
redshift bins. \label{fig6}}
\end{figure}

Figure \ref{fig6} shows photometric magnitude differences of quasars between 
the SCUSS and SDSS $u$ bands as a function of redshift. The 
photometric errors are limited to 0.1 mag, which approximately 
corresponds to the quasar redshift up to about 3. We find that the 
average quasar variability varies at a level of 0.04 mag in the 
redshift range of 0--3. Thus, the two-epoch variability is almost 
independent on the redshift, which can help to discover more quasars 
with $0 < z < 3.5$ where variability is still available. The independence 
of variability on redshift in $gri$ bands was also confirmed by \citet{zuo12}
who divided the quasars into different subsamples with different physical 
parameters.  

\section{Quasar Selection and Analysis} \label{sec4}
\subsection{Ignorable variability}
Quasar target selections in this section are investigated based on the 
consideration of whether objects presenting variability. Those objects 
with two-epoch variability less than three times the photometric errors 
are regarded as sources with ignorable or small variabilities, which can 
be expressed as
\begin{equation}
\Delta m = |u_\mathrm{SCUSS} - u_\mathrm{SDSS}| \leq 3\sigma,
\end{equation}
where $\sigma = \sqrt{\sigma_{u_\mathrm{SCUSS}}^2+\sigma_{u_\mathrm{SDSS}}^2}$ 
and $\sigma_{u_\mathrm{SCUSS}}$ and $\sigma_{u_\mathrm{SDSS}}$ are the SCUSS 
and SDSS $u$-band magnitude errors, respectively. There are 68.4\% quasars 
and 96.9\% stars of their total samples with ${\Delta}m \leq 3\sigma$. These 
objects have deeper SCUSS $u$ band and hence deeper intrinsic colors. It is 
imaginable that the SCUSS $u$ band instead of the SDSS $u$ band would 
undoubtedly improve the efficiency of the quasar selection when combined 
with SDSS other bands. 

In order to show the advantage of the quasar selection based on SCUSS data, 
we introduce a flux-based quasar target selection algorithm, XDQSO, which 
applies the extreme-deconvolution method to evaluate the probability that 
an object is a quasar \citep{bov11}. By applying the XDQSO method to 
the above non-variable samples, we can know how many more quasars can be 
selected with SCUSS $u$ and SDSS $griz$ data than with SDSS-only data. The 
probability of an object that selected as a quasar in XDQSO is set to be 
larger than 0.9. 

\begin{figure} \epsscale{1.0} \plotone{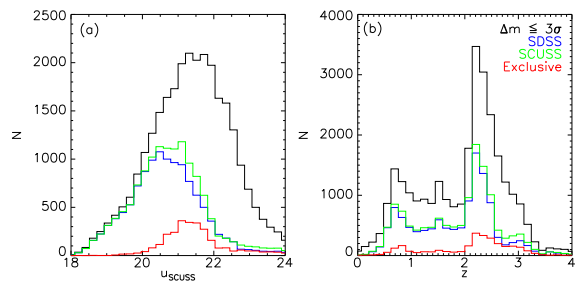}
\caption{Magnitude (a) and redshift (b) distributions of quasars selected by the 
XDQSO method respectively applied to SCUSS $u$ plus SDSS $griz$ data (green) and 
SDSS-only data (blue). Distributions of the total quasar samples with ignorable variability 
(black) and quasars exclusively selected by XDQSO with SCUSS plus SDSS data (red) are 
also overlapped.
\label{fig7}}
\end{figure}

After the XDQSO method is applied to the non-variable samples with SCUSS $u$ 
and SDSS $griz$ photometric data, about 46.6\% quasars and 2.9\% stars are 
selected. When it is applied to SDSS $ugriz$ data, about 41.4\% quasars and 
2.4\% stars are selected. There are about 24.0\% quasars that are exclusively 
selected by XDQSO with SCUSS $u$ and SDSS $griz$ data. These numbers are 
summarized in Table \ref{tab1}. Figure \ref{fig7} shows the magnitude and 
redshift distributions of quasars selected by these two methods. The 
distributions of total samples with ignorable variability and quasars 
exclusively selected by the XDQSO with SCUSS data are also overlaid in this 
figure. The target selection with SCUSS data can select more faint quasars 
with magnitude peak at 21.5 mag and more quasars with redshift between 2 and 
3 (peak at $z = 2.2$). We also check the quasar selection efficiencies 
within two different redshift ranges: $0< z <2$ and $2< z < 3.5$, which are 
also presented in Table \ref{tab1}. As a result, there are respectively 
15.6\% and 31.2\% more quasars that are selected by the XDQSO with SCUSS 
data than that with SDSS-only data. The selection here is based on known 
quasar samples, most of which are selected through SDSS optical colors. The 
SCUSS $u$ band is deeper than that of the SDSS, which implies that a part of 
potential quasars close to the SDSS $u$-band magnitude limit are missing. The 
SCUSS data can help us find more faint quasars than the above experiment only 
based on the known samples.

\begin{table*}
\centering
\caption{Quasar selection for objects with ignorable variability \label{tab1}}
\begin{tabular}{c|c|c|c|c}
\hline
\hline
 & Star & Quasar & $0 < z < 2$ & $2 < z < 3.5$ \\
\hline
Total & 56518 & 42418  & 20074 & 21366 \\
${\Delta}m \leq 3\sigma$ & 54744 (96.9\%) & 29011 (68.4\%) & 11458 (57.1\%) & 16610 (77.7\%) \\
\hline
SCUSS $u$ + SDSS $griz$ & 1570 (2.9\%) & 13511 (46.6\%) & 5998 (52.3\%) & 7274 (43.8\%) \\
SDSS $ugriz$ & 1335 (2.4\%) & 12024 (41.4\%) & 5548 (48.4\%) & 6266 (37.7\%) \\
\hline
Exclusive & \nodata & 2884 (24.0\%) & 865 (15.6\%) & 1958 (31.2\%) \\
\hline
\end{tabular}
\tablecomments{The 2nd and 3rd columns show the selection efficiencies with the total 
star and quasar samples. The 4th and 5th columns show the quasar selection efficiencies 
within two different redshift ranges. The 2nd row gives sample numbers. The 3rd row 
gives corresponding samples with ignorable variability (${\Delta}m \leq 3\sigma$). 
The 3rd and 4th rows respectively present the selection efficiencies of two target selections: 
XDQSO with SCUSS $u$ plus SDSS $griz$ data and XDQSO with SDSS-only data. The 
last row gives the quasars exclusively selected by XDQSO with SCUSS $u$ and SDSS $griz$ 
data relative to XDQSO with SDSS-only data. }
\end{table*}

\subsection{Obvious variability}
\begin{figure} \epsscale{1.0} \plotone{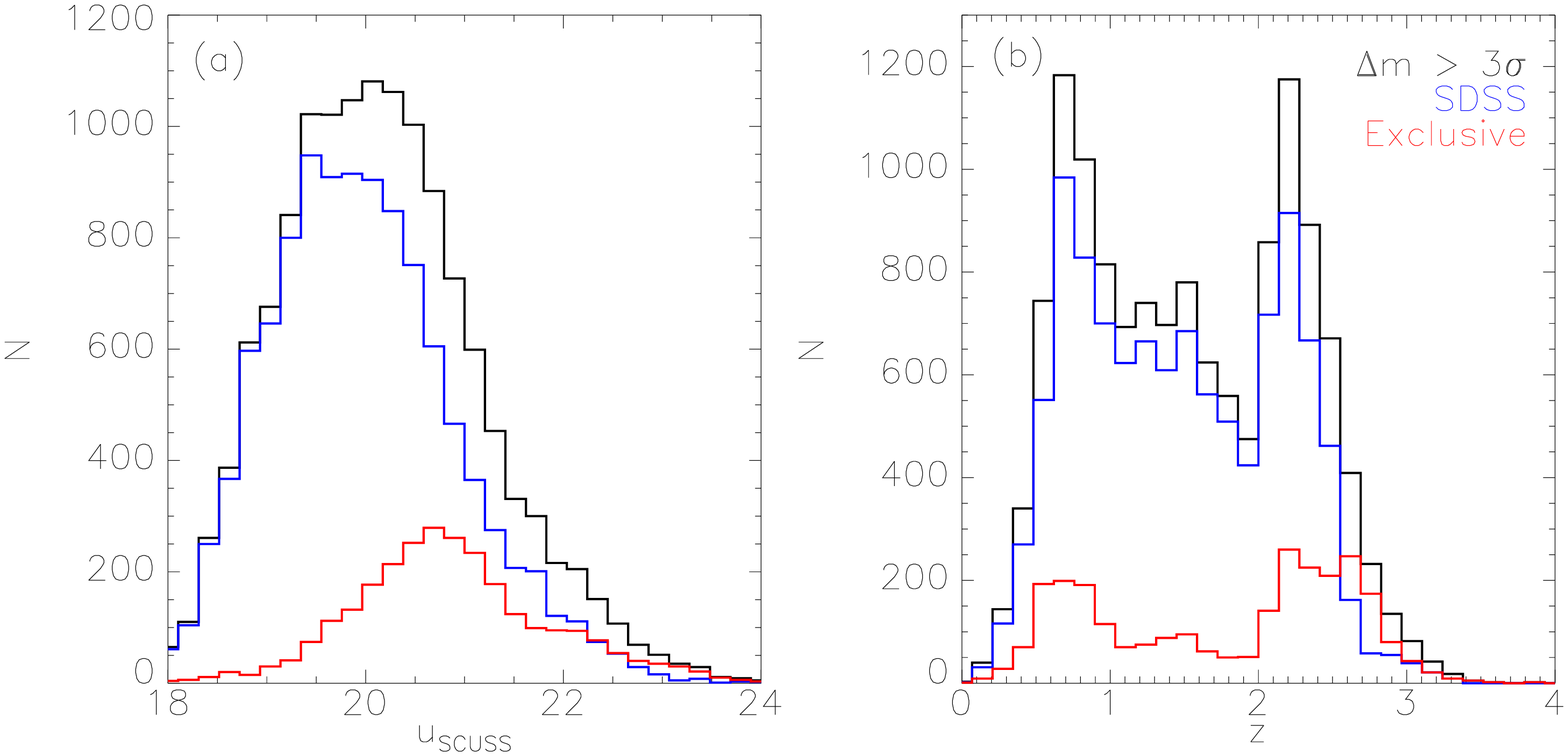}
\caption{Magnitude (a) and redshift (b) distributions of quasars selected by 
the two-epoch variability (black) and XDQSO method with SDSS-only data (blue). 
Distributions of the total quasar samples with obvious variability 
(black) and quasars exclusively selected by variability (red) are also 
plotted. \label{fig8}}
\end{figure}

Objects with ${\Delta}m > 3\sigma$ are considered as sources with obvious 
or large variability. About 31.6\% quasars and only 3.1\% stars have obvious 
two-epoch variability. The fraction of stars having large variability is 
much smaller than that of quasars, so the objects with ${\Delta}m > 3\sigma$ 
is regarded as quasar candidates. We compared the selection by the two-epoch 
variability with that by the XDQSO method applied to the SDSS $ugriz$ data 
of large-variability samples. About 80\% quasars and  11.2\% stars are 
selected by XDQSO. There are 25.7\% quasars exclusively selected by 
variability. These numbers are summarized in Table \ref{tab2}. Figure 
\ref{fig8} shows the magnitude and redshift distributions of quasars obtained 
by these two selections. The selection by two-epoch variability can 
select more faint quasars with magnitude peak at 20.5 mag. It 
also can select more low-redshift quasars with ultraviolet excess (peak 
at $z = 0.7$) and more mid-redshift quasars with $2< z < 3$ (peak at 
$z = 2.5$). There are 17.3\% and 43.3\% quasars exclusively selected by 
variability for the two redshift ranges as specified in the previous 
section, which are also shown in Table \ref{tab2}. 

\begin{table*}
\centering
\caption{Quasar selection for objects with obvious variability \label{tab2}}
\begin{tabular}{c|c|c|c|c}
\hline
\hline
 & Star & Quasar & $0 < z < 2$ & $2 < z < 3.5$ \\
\hline
Total & 56518 & 42418  & 20074 & 21366 \\
${\Delta}m > 3\sigma$ & 1774 (3.1\%) & 13407 (31.6\%) & 8616 (42.9\%) & 4756 (22.3\%) \\
\hline
SDSS $ugriz$ & 199 (11.2\%) & 10669 (79.6\%) & 7346 (85.3\%) & 3318 (69.8\%) \\
\hline
Exclusive & \nodata & 2738 (25.7\%) & 1270 (17.3\%) & 1438 (43.3\%) \\
\hline
\end{tabular}
\tablecomments{Similar to Table \ref{tab1}, but for quasars with large variability. 
Objects with the two-epoch variability ${\Delta}m > 3\sigma$ are considered as quasar 
targets. The last row gives the quasars exclusively selected by variability relative 
to the selection by XDQSO with SDSS-only data. }
\end{table*}

The SDSS $u$-band photometric error is larger than the SCUSS $u$ band, which is 
more prominent close to the SDSS magnitude limit. So the selection by 
variability is limited to a brighter magnitude range than the XDQSO selection 
with SCUSS $u$ plus SDSS $griz$ data. The XDQSO selection with the  
SCUSS plus SDSS data can select more fainter quasars and quasars with 
redshift peaked at 2.2, while the selection by variability can select more 
low-redshift and mid-redshift quasars. These two selections are complementary 
and make the most of the SCUSS $u$-band advantages. 

\subsection{Application of our selection}
We randomly select a region of about 10 deg$^2$ with consistent SCUSS 
imaging quality to demonstrate the capability of our quasar selection with 
full point sources from the SDSS. The quasar probability in XDQSO is still set 
to be larger than 0.9. The SCUSS u-band apparent magnitude is limitted to 23.5 
mag. The $i$-band magnitude is between 18 and 22.5 mag. Isolated sources 
classified as point-like by the SDSS are considered. When objects present 
ignorable variability, the selection with SCUSS and SDSS data can find 29\% 
more quasar targets than the one only with the SDSS data. On the other hand, 
the selection by variability can find 168.7\% additional targets, although 
star comtamination would be more serious. In total, we find that our selection 
can select 87.7\% additional quasar targets. However, its actual application 
with alterable parameters, such as the quasar probability 
in the XDQSO and the variability amplitude (several times larger than the 
photometric error), should be dependent on the requirements of the density 
and homogeneity of quasar targets, the selection efficiency, and magnitude 
range, etc.

\section{Conclusion} \label{sec5}
The SCUSS provides a deep and wide imaging survey in an about 5000 
deg$^2$ area of the south Galactic cap with $u$ band. The SCUSS $u$-band is 
deeper than the SDSS $u$ band and the observation time lag between the 
SCUSS and SDSS provides an opportunity to investigate the two-epoch 
variability. Through combining the SCUSS and SDSS data, we evaluate the 
capability of the quasar selection in consideration of the advantages 
of the SCUSS $u$ band data. 

The deeper SCUSS $u$-band photometry makes stars and quasars more tightly 
distributed in the color-color diagrams if SDSS other bands are included 
and objects present no variability. It helps to select more fainter quasars. 
Quasars have power-law continua and many strong emission lines. The photometric
difference for quasars due to different photometric systems is larger than 
that of stars, which is helpful for the separation of these two 
objects. The average variability of quasars is about 0.3 mag. The 
distribution of the two-epoch variability between the SCUSS and SDSS for 
quasars is quite different from that of stars. Besides, the quasar two-epoch 
variability is independent on redshift at a level of 0.04 mag ($0 < z < 3$).

Based on the above advantages, we analyze the quasar selection in two ways: 
one is combining the SCUSS deeper $u$-band data and SDSS $griz$ data 
when objects present ignorable variability; the other is utilizing the 
two-epoch variability. The XDQSO method, which gives the quasar probability, 
is introduced for comparisons. We find that the XDQSO method with SCUSS $u$ 
and SDSS $griz$ data can select more faint quasars and quasars with redshift 
around 2.2 than the XDQSO with SDSS $ugriz$ data. There are about 24.0\% 
quasars exclusively selected by the former. The SCUSS data can also help 
us find the missing quasars close to the SDSS $u$-band magnitude limit. 
We regard objects with ${\Delta}m > 3\sigma$ as quasar candidates 
with obvious variability, because there are relatively small fraction of 
stars with such large variability. The quasar selection by variability can 
select both low-redshift quasars and mid-redshift quasars with $2< z < 3$. 
There are 25.7\% quasars exclusively selected by variability. The 
above two quasar selections are complementary, which make full use of the 
SCUSS advantages. When applying to the full SDSS point sources, our method 
can also select about 87.7\% additional quasar targets.

\acknowledgments
This work is supported by the National Natural Science Foundation of China (NSFC, Nos. 11203031, 11433005, 11073032, 11373035, 11203034, 11303038, 11303043, 11033001, 11373008), by the National Basic Research Program of China (973 Program, Nos. 2014CB845704, 2013CB834902, 2014CB845702, and 2014CB845705), and by the Strategic Priority Research Program "The Emergence of Cosmological Structures" of the Chinese Academy of Sciences(No. XDB09000000). Z. Y. Wu was supported by the Chinese National Natural Science Foundation grant No. 11373033. This work was also supported by the joint fund of Astronomy of the National Nature Science Foundation of China and the Chinese Academy of Science, under Grants U1231113. The SCUSS is funded by the Main Direction Program of Knowledge Innovation of Chinese Academy of Sciences (No. KJCX2-EW-T06). It is also an international cooperative project between National Astronomical Observatories, Chinese Academy of Sciences, and Steward Observatory, University of Arizona, USA. Technical supports and observational assistances of the Bok telescope are provided by Steward Observatory. The project is managed by the National Astronomical Observatory of China and Shanghai Astronomical Observatory.

SDSS-III is managed by the Astrophysical Research Consortium for the Participating Institutions of the SDSS-III Collaboration including the University of Arizona, the Brazilian Participation Group, Brookhaven National Laboratory, Carnegie Mellon University, University of Florida, the French Participation Group, the German Participation Group, Harvard University, the Instituto de Astrofisica de Canarias, the Michigan State/Notre Dame/JINA Participation Group, Johns Hopkins University, Lawrence Berkeley National Laboratory, Max Planck Institute for Astrophysics, Max Planck Institute for Extraterrestrial Physics, New Mexico State University, New York University, Ohio State University, Pennsylvania State University, University of Portsmouth, Princeton University, the Spanish Participation Group, University of Tokyo, University of Utah, Vanderbilt University, University of Virginia, University of Washington, and Yale University.

\end{document}